\documentclass[%
 reprint,
superscriptaddress,
nobibnotes,
 amsmath,amssymb,
 aps,
prb,
floatfix,
]{revtex4-2}

\usepackage[utf8]{inputenc}
\usepackage{graphicx}
\usepackage{dcolumn}
\usepackage{bm}
\usepackage{xcolor}
\usepackage{microtype}
\usepackage[
    separate-uncertainty = true,
    locale=US,
    ]{siunitx}
\usepackage{cleveref}

\begin{document}

\title{Pathways to Bubble and Skyrmion Lattice Formation in Fe/Gd Multilayers} 

\author{Tim Titze}
\affiliation{I.\,Physikalisches Institut, Universit\"at G\"ottingen, 37077 G\"ottingen, Germany\looseness=-1}
\author{Sabri Koraltan}
\affiliation{Institute of Applied Physics, Technische Universität Wien, A-1040, Vienna, Austria}
\affiliation{Physics of Functional Materials, Faculty of Physics, University of Vienna, Vienna, Austria\looseness=-1}
\affiliation{Vienna Doctoral School in Physics, University of Vienna, Vienna, Austria}
\author{Mailin Matthies}
\affiliation{I.\,Physikalisches Institut, Universit\"at G\"ottingen, 37077 G\"ottingen, Germany\looseness=-1}
\author{Timo Schmidt}
\affiliation{Institute of Physics, University of Augsburg, 86135 Augsburg, Germany\looseness=-1}
\author{Dieter Suess}
\affiliation{Physics of Functional Materials, Faculty of Physics, University of Vienna, Vienna, Austria\looseness=-1} 
\author{Manfred Albrecht}
\affiliation{Institute of Physics, University of Augsburg, 86135 Augsburg, Germany\looseness=-1}
\author{Stefan Mathias}
\email[]{smathias@uni-goettingen.de}
\affiliation{I.\,Physikalisches Institut, Universit\"at G\"ottingen, 37077 G\"ottingen, Germany\looseness=-1}
\affiliation{International Center for Advanced Studies of Energy Conversion (ICASEC), Universit\"at G\"ottingen, 37077 G\"ottingen, Germany} 
\author{Daniel Steil}
\email[]{dsteil@gwdg.de}
\affiliation{I.\,Physikalisches Institut, Universit\"at G\"ottingen, 37077 G\"ottingen, Germany\looseness=-1}

\begin{abstract}
The creation and control of magnetic spin textures is of great interest in fundamental research and future device-oriented applications. Fe/Gd multilayers host a rich variety of magnetic textures including topologically trivial bubbles and topologically protected skyrmions. Using time-resolved Kerr spectroscopy, we highlight how various control strategies, including temperature, out-of-plane magnetic fields and femtosecond light excitation, can be used to create such textures via different pathways. We find that varying the magnetic field for constant temperature leads to a different ($H, T$)-phase diagram of magnetic textures than moving along a temperature trajectory for constant magnetic field. Micromagnetic simulations corroborate this finding and allow to visualize the different paths taken. Furthermore, we show that the creation of bubbles and skyrmions in this material via impulsive light excitation is not solely governed by temperature-driven processes, since bubbles and skyrmions can be stabilized in parts of the ($H, T$)-phase diagram, where neither the constant temperature nor the constant magnetic field trajectory predict their existence. Using this phase diagram, we reason why bubble and skyrmion creation in this particular system is only possible from the stripe domain state. Our observations provide a versatile toolkit for tailoring the creation of magnetic spin textures in Fe/Gd multilayers.
\end{abstract}

\maketitle
\section{Introduction}
Deterministic control of magnetic spin textures promises new avenues for future spintronic and magnonic devices~\cite{Chumak2015,Finocchio2016,Fert2017,Barman2021,Yu2021,Chumak2022,Petti2022}. In particular, topologically nontrivial magnetic skyrmions are possible key drivers for innovative applications including unconventional computing concepts~\cite{Huang2017,Song2020,Li2021,Yokouchi2022}. However, the possible pathways to realize different spin textures in skyrmion host materials are still not fully explored. Studies in this regard include, e.g., Refs.~\cite{Montoya2017a, Montoya2017b} and typically rely on imaging techniques like Lorentz transmission electron microscopy (LTEM)~\cite{Grundy1968,Petford-Long2012, McCray2021} or magnetic X-ray imaging~\cite{Fischer1998, Fischer2015} for the detection of different magnetic spin textures. In the last years, time domain techniques like optical Kerr spectroscopy have, however, proven to be effective in characterizing the magnetic spin textures of thin film sample system via the coherent response of the spin system to ultrafast optical stimulation~\cite{Ogawa2015,Padmanabhan2019,Sekiguchi2022,Kalin2022,Titze2024a,Kalin2024}.

We have recently utilized Kerr spectroscopy in conjunction with LTEM studies and micromagnetic simulations to demonstrate the detection and transformation of different ground state magnetic spin textures in a [Fe(0.35 nm)/Gd(0.40 nm)]$_{160}$ multilayer system~\cite{Titze2024a}. We further demonstrated active control of the coherent dynamics of bubbles and skyrmions in this system~\cite{Titze2024b}. 

\begin{figure}[htb]
     \centering
     \includegraphics[width=\columnwidth]{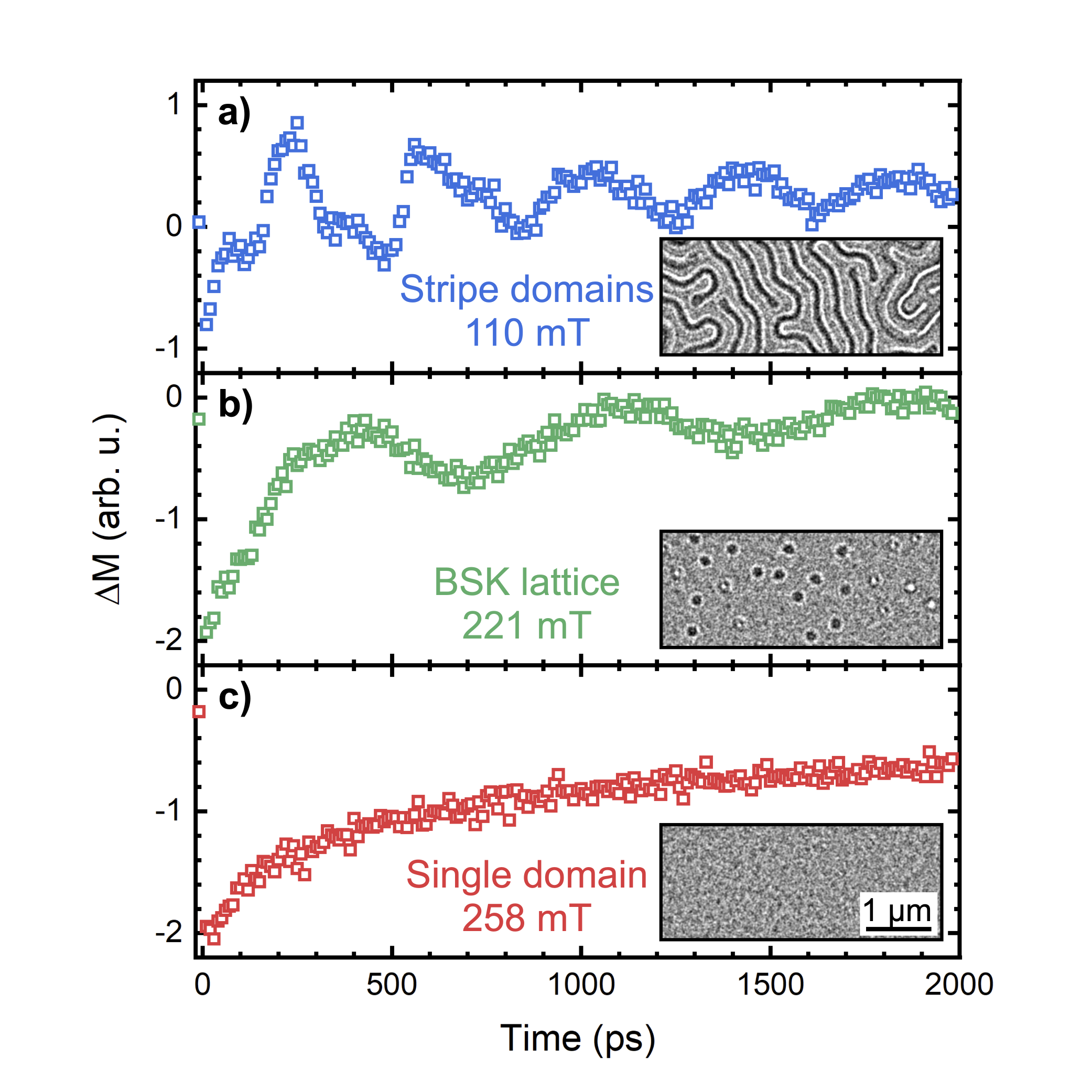}
     \caption{\textbf{Magnetization dynamics and corresponding magnetic spin textures.} Laser-induced magnetization dynamics for \textbf{a)} stripe domains, \textbf{b)} the BSK lattice, and \textbf{c)} the single domain phase with corresponding LTEM images of the magnetic spin texture as an inset.}
     \label{fig:Overview}
\end{figure}

Figure~\ref{fig:Overview} shortly summarizes the laser-induced magnetization dynamics as well as ground state magnetic spin textures observed in this material system for different applied out-of-plane (oop) magnetic fields. For details we refer to Ref.~\cite{Titze2024a}. Evidently, three inherently different phases of magnetic texture are observed, including stripe domains, a bubble and skyrmion (BSK) lattice and a single domain state depending on the external oop magnetic field. We note in passing, that unlike in typical skyrmion host materials, the creation of skyrmions in this material system is not driven by the Dzyaloshinskii–Moriya interaction (DMI). Instead, it arises from a competition among dipolar interactions, perpendicular magnetic anisotropy, and exchange interactions~\cite{Montoya2017a, Heigl2021}. The different magnetic textures are identified by unique signatures in Kerr spectroscopy, corresponding to distinct coherent spin wave modes of stripe domains $m_{\mathrm{st}}$ and BSKs $m_{\mathrm{bsk}}$ on the nanosecond timescale. In contrast, only an incoherent magnetization recovery is observed for the single domain state at ambient temperature, as reported in Ref.~\cite{Titze2024a}.  

In this work, we aim to extend on Ref.~\cite{Titze2024a} by (i) studying the role of sample base temperature and external magnetic field on the observed magnetic spin texture for weak optical perturbation of the sample. We both sweep the oop magnetic field $H$ for constant $T$, and also sweep $T$ for constant $H$ and measure the resulting coherent spin dynamics using Kerr spectroscopy. From this data we create ($H, T$) spin texture phase diagrams showing the temperature and magnetic field parameters required to stabilize each texture. In particular, we find that the presence of the BSK lattice state strongly depends on the path travelled, which is corroborated by finite-temperature micromagnetic simulations using magnum.np~\cite{Bruckner2023}, where thermal fluctuations are included with a white noise of vanishing mean value. Further, (ii) we study the differences in phase diagrams for weak and strong optical excitation, i.e., for laser-fluences either purely exciting or additionally transforming the ground state spin textures, as previously shown at ambient $T$ in~\cite{Titze2024a}. Here, we find that the transformation of spin textures cannot simply be explained in terms of a pure temperature increase by the laser-excitation and subsequent quench of a high-$T$ adiabatically accessible spin texture into a metastable state. Using this data, we propose a model for BSK nucleation in this material system.
 
\section{Methods}
Multilayers were grown by DC magnetron sputtering at room temperature on thermally oxidized Si(100) substrates as well as on Si$_3$N$_4$ membranes required for LTEM measurements. The working gas was argon and the vacuum base pressure was below $1\times10^{-8}$\,mbar. After depositing a 5\,nm Pt seed layer, the [Fe(0.35\,nm)/Gd(0.40\,nm)]$_{160}$ multilayers were sputtered at an Ar pressure of 3.5\,{\textmu}bar. Finally, the layer stack was capped with a $5$\,nm Si$_3$N$_4$ layer using RF sputtering at 1.5\,{\textmu}bar with nitrogen gas in the chamber. 

Time-resolved data is recorded by a pump-probe magneto-optical Kerr effect setup using a fiber laser amplifier system at 1030\,nm central wavelength with subsequent spectral broadening in a gas-filled hollow core fiber. Pulse compression is performed using a pair of chirped mirrors yielding laser pulses of $<40$\,fs pulse duration. The linearly polarized fundamental laser pulses at a repetition rate of 50\,kHz are split into two identical replica using a thin beam splitter and the replica used as a probe pulse is frequency doubled using a BBO crystal to avoid bleaching effects~\cite{Koopmans2000, Razdolski2017} and to allow for easy separation of pump and probe by a combination of dichroic mirrors and color filters in front of the detector. A separate pulse compression scheme for the probe is used to obtain a laser pulse duration of less than 40\,fs in front of the sample. The pump-probe delay is tuned using a 2\,ns linear delay stage in the pump arm of the setup and the pump beam is additionally chopped, allowing to detect both the pumped and unpumped signal from the sample. Magnetic field and temperature control is achieved by utilizing a combination of variable gap electromagnet and liquid helium finger cryostat with optical access. After reflection from the sample close to normal incidence the probe polarization is analyzed using a balanced bridge detector connected to a 250 MHz digitizer card with 16bit vertical resolution, which allows to measure the amplified difference signal corresponding to the observed polar Kerr rotation.

\section{Results: Weak perturbation}
\subsection{Data acquisition and analysis}
Measurements in the weak perturbation regime were taken at an incident pump fluence of $F\approx 0.5$\,mJ/cm$^2$. The measured signal consists of incoherent demagnetization on the sub-ps timescale and a magnetization recovery (two-step, picosecond and nanosecond timescale), which is superposed by a coherent signal contribution on the nanosecond timescale in case magnetic spin textures are present (see Fig.~\ref{fig:Overview} for a representative curve). To analyze the coherent signal contribution, we first subtract the incoherent signal contribution from the data and start the data analysis not directly at time zero (for details see~\cite{Titze2024a}). We then employ a fast Fourier transform (FFT) to obtain the frequency of the breathing modes for three different $(H,T)$-pathways: (i) constant temperature, increasing magnetic field; (ii) constant field, increasing temperature and (iii) constant field, decreasing temperature, see Fig.~\ref{fig:BandTsweep}a,\,c,\,e. The case of constant temperature and decreasing magnetic field was not studied, as it is already known from Ref.~\cite{Titze2024a} that no nucleation of a BSK lattice takes place by decreasing the magnetic field from saturation.

\subsection{Weak perturbation - phase maps of spin textures}
\begin{figure*}[t]
     \centering     \includegraphics[width=\textwidth]{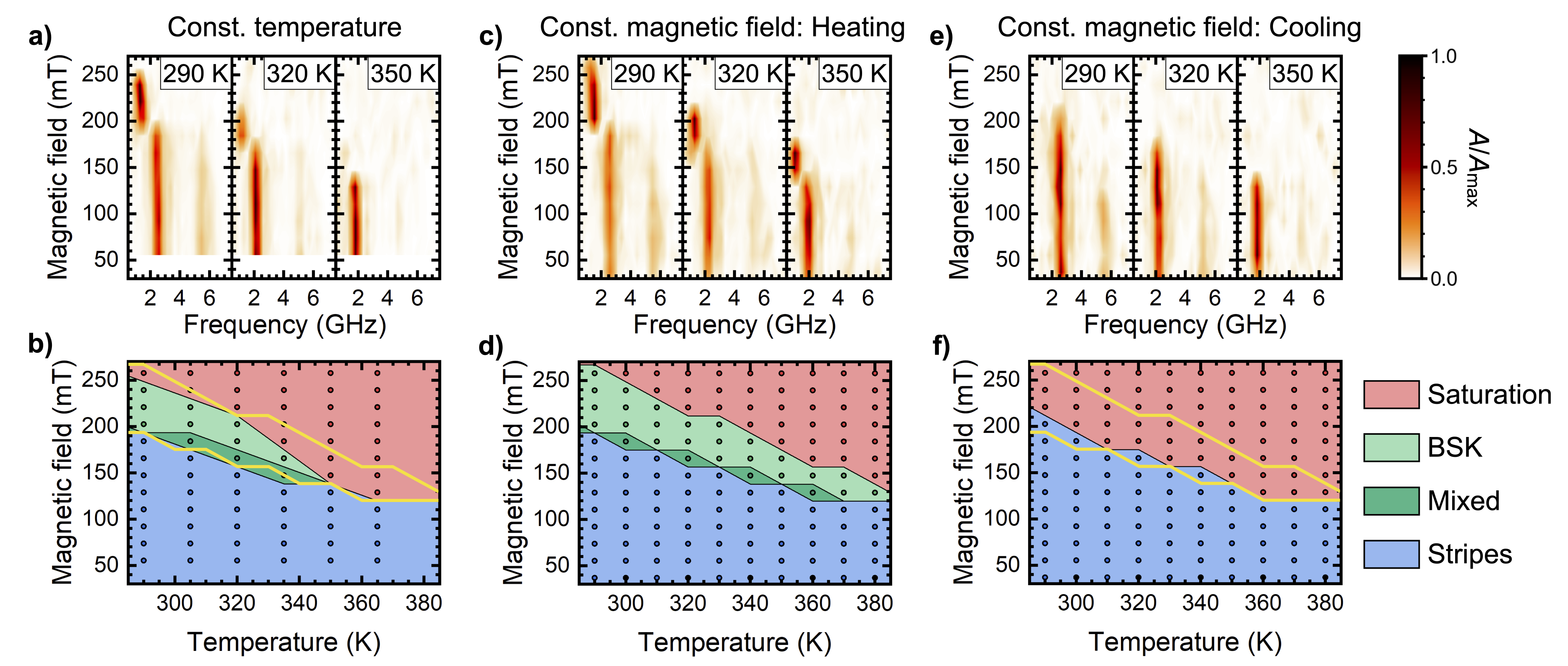}
     \caption{\textbf{Magnetic field- and temperature-dependent magnetic spin textures.} Spin-wave modes at different $T$ extracted from Fourier analysis of the MOKE data for measurements at \textbf{a)} constant temperature and increasing magnetic field, \textbf{c)} heating at constant magnetic field, \textbf{e)} cooling at constant magnetic field. Fourier amplitudes are normalized in all figures. Phase diagram for weak optical excitation obtained for \textbf{b)} increasing the magnetic field at constant sample temperature, \textbf{d)} heating at constant magnetic field, \textbf{f)} cooling at constant magnetic field. The colors denote the different spin textures: stripe domains (blue), BSK lattice (green) and magnetic saturation (red). Dark green color denotes transition regions between the stripe domain and BSK textures. The yellow outlines in b) and f) highlight the size of the BSK region  from d). Data points for a given ($H,T$) correspond to the TR-MOKE measurements. Concerning d) and f), the magnetic state is reset before each temperature sweep by applying a negative saturating magnetic field.}
     \label{fig:BandTsweep}
\end{figure*}

In all cases, the FFT yields clear and well separated mode frequencies $f_{\mathrm{st}}$ for the stripe domain state and $f_{\mathrm{bsk}}$ for the BSK lattice state with barely any dispersion in the magnetic field, see Fig.~\ref{fig:BandTsweep}a,\,c,\,e. However, temperature does significantly influence the mode frequencies, as shown in the Appendix, Fig.~\ref{fig:Amplitudes}. 

The presence of each phase strongly depends on the oop magnetic field and temperature, which can be expressed using ($H,\,T$)-phase diagrams, see Fig.~\ref{fig:BandTsweep}b,\,d,\,f discussed below.  

\subsubsection{Increasing magnetic field at constant sample temperature}

Figure~\ref{fig:BandTsweep}b depicts the ($H,T$)-phase diagram for measurements at constant temperature and increasing magnetic field. Three regions of magnetic texture are clearly visible. At magnetic fields $\mu_0 H\leq110$\,mT the sample is in a stripe domain state for all $T$ (blue color). This range gradually extends up to about $\mu_0 H=190$\,mT for decreasing T. At $\mu_0 H\geq 250$\,mT the sample is in a saturated state for all studied $T$ (red color). For increasing $T$, the saturated state is reached for lower and lower $T$ down to about $\mu_0 H=120$\,mT. In between the stripe domain and saturated state, we find a pocket for $T\leq 340$\,K, where a BSK lattice is observed (green colors). This region shifts from $\mu_0 H\approx250$\,mT down to $\mu_0 H\approx150$\,mT field with increasing T, until it finally vanishes at $T\approx350$\,K and the sample directly transitions from the stripe domain state into the saturated magnetic state. Therefore, the stability ranges of the stripe and BSK phases narrow with increasing $T$, as the saturated magnetic state is reached for smaller and smaller $H$.

\subsubsection{Increasing sample temperature at constant magnetic field}
Figure~\ref{fig:BandTsweep}d depicts the ($H,T$)-phase diagram in case of temperature increase at constant magnetic field. Again, three regions of magnetic texture can be clearly distinguished and, similar to before, the saturated magnetic state occurs for lower and lower magnetic fields for increasing $T$. In contrast to before, however, even for the highest measured temperatures $T\geq380$\,K a BSK phase is present in the phase diagram, i.e., the stability range of BSKs is drastically increased. This difference to the phase diagram depicted in Fig.~\ref{fig:BandTsweep}b illustrates that the presence of each spin texture phase does not only depend on the combination of magnetic field and temperature, but also on the path travelled through the ($H,T$)-phase diagram. 

To further emphasize this difference, we have highlighted the region where the BSK lattice appears in Fig.~\ref{fig:BandTsweep}d by outlining it in yellow in Fig.~\ref{fig:BandTsweep}b. For temperatures $T<320$\,K the positions of the BSK lattice phase matches well for both pathways, although the BSK lattice phase seems to be slightly broader if the temperature is increased at constant magnetic field. In particular, the transition from the BSK lattice to the single domain state is affected, i.e., higher magnetic fields are evidently required to annihilate BSKs. This effect is even more prominent considering temperatures $320$\,K$<T<350$\,K since the width of the BSK phase remains almost constant in Fig.~\ref{fig:BandTsweep}d, while in Fig.~\ref{fig:BandTsweep}b the saturated state is already reached, as indicated by the red region. As a result, we find a BSK lattice stabilized at high temperatures of at least $380$\,K in Fig.~\ref{fig:BandTsweep}d that is only accessible via heating at constant magnetic field. 

\subsubsection{Decreasing sample temperature at constant magnetic field}
Last, keeping the magnetic field $H=\textrm{const.}$ and decreasing $T$, as shown in Fig.~\ref{fig:BandTsweep}f does not lead to any BSK formation. A direct transition occurs from the single domain state to the stripe domain state. Again, the region outlined in yellow denotes the BSK lattice phase from Fig.~\ref{fig:BandTsweep}d. These results highlight that the saturated domain state now extends down in field to the border of the stripe domain state from Fig.~\ref{fig:BandTsweep}d. The absence of BSKs in favor of a single domain state is comparable to the downsweep of the magnetic field at constant temperature reported in~\cite{Titze2024a}. Thereby, it is corroborated that adiabatic BSK nucleation in this sample system requires the presence of stripe domains.  

\subsection{Micromagnetic simulations}
\begin{figure}[tb]
     \centering
     \includegraphics[width=\columnwidth]{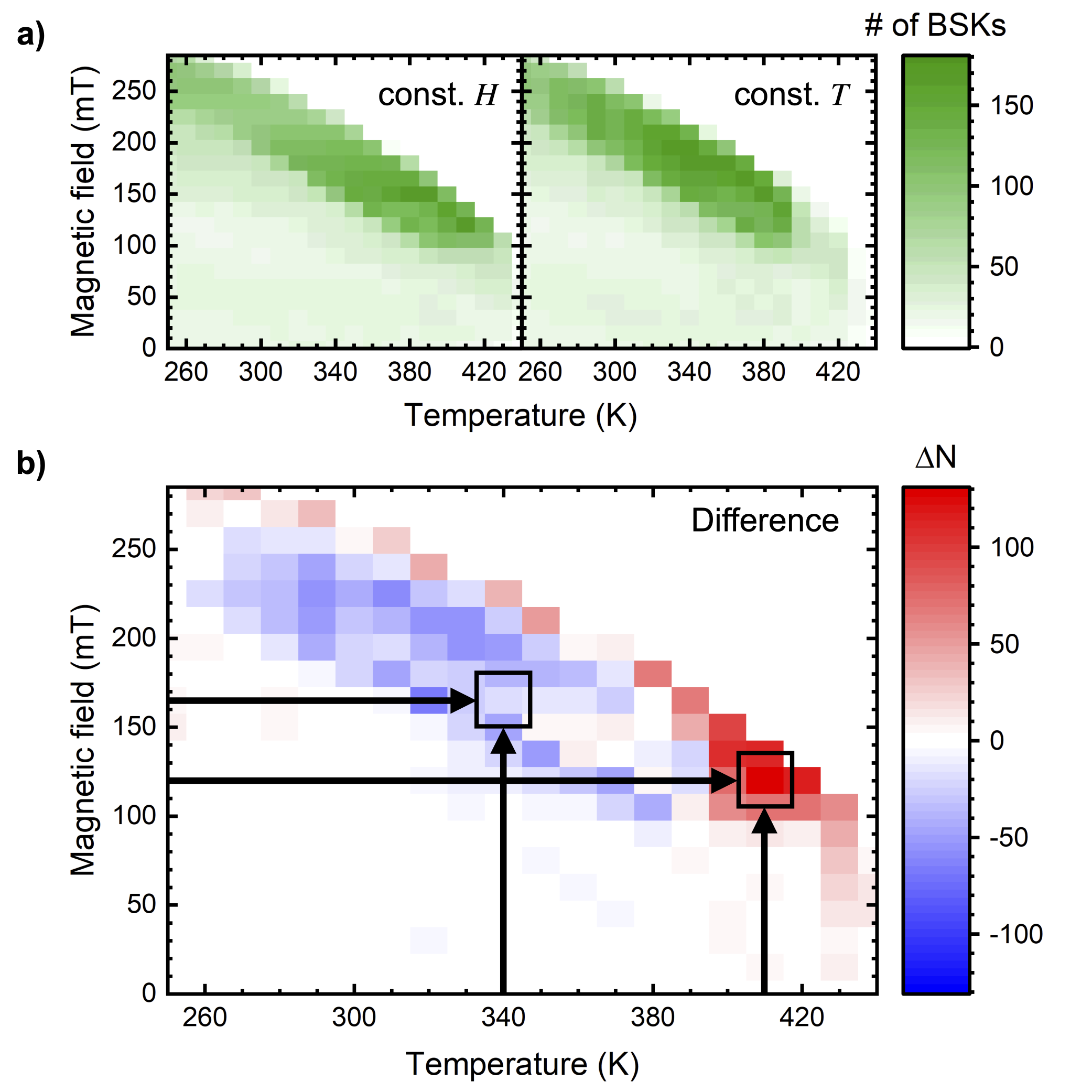}
     \caption{\textbf{Magnetic field- and temperature-dependent phase diagrams obtained from micromagnetic simulations.} \textbf{a)} Number of BSKs within the $5.12\times 5.12$\,{\textmu}m$^2$ simulated area for heating at constant magnetic field and increasing the magnetic field at constant sample temperature. \textbf{b)} Difference of the data shown in \textbf{a)} revealing pathway-dependent magnetic spin textures. Arrows mark different pathways toward selected ($H, T$) discussed in the text.}
     \label{fig:Sim}
\end{figure}

To obtain a microscopic view on the different phase diagrams for increasing $H$ at constant $T$ and increasing $T$ at constant $H$, we performed micromagnetic simulations of the Fe/Gd multilayer stack with material parameters identical to previous work~\cite{Titze2024a}, but including a variable sample temperature. We simulate a box with ($512\times 512\times 10$) cells with a cell discretization of $(\SI{10}{nm}, \SI{10}{nm}, \SI{12}{nm})$. Sample parameters are room temperature are $M_\textrm{S} = \SI{340}{kA/m}$, $K_\textrm{U} = \SI{40}{kJ/m^3}$ and $A = \SI{6}{pJ/m}$. The temperature dependencies are calculated as $
M_{\mathrm{sat}}(T) = M_{\mathrm{sat}}(\SI{0}{K}) \cdot [1 - (T/T_c)^2]$,
$K_{\mathrm{U}} = K_{\mathrm{U}}(0\,\mathrm{K}) \cdot [M_\textrm{S}(T)/M_\textrm{S}(\SI{0}{K})]^{3}$ and $
A(T) = A(0\,\mathrm{K}) \cdot [M_\textrm{S}(T)/M_\textrm{S}(\SI{0}{K})]^{1.7}$. 
For details see Ref.~\cite{Titze2024b}. These parameters allow for a qualitative comparison with the experimental data. To reproduce the experimental protocol, we generate initial magnetization configurations at $T=250$\,K and vanishing external fields. For this, we start from a random magnetic state with a uniform distribution. We relax into an energetic minimum by solving the Landau-Lifshitz-Gilbert equation for at least 10\,ns at low damping $\alpha = 0.02$, including thermal fluctuations~\cite{Leliaert2017}. We now separate the simulations performed for constant $H$ and increasing $T$, and constant $T$ and increasing $H$. For the former, we use the state at $T = 250$\,K, and increase the magnetic field in steps of $15$\,mT. Each state is relaxed for another $10$\,ns. Then, we use these as the new initial states, and for each constant field value, we increase the temperature from $250$\,K to $450$\,K in steps of $10$\,K with a relaxation time of $10$\,ns. For the case of constant $T$ and increasing $H$, we follow an analogous approach, where we first increase the temperature starting from the state with $T = 250$\,K at zero field. With the new initial magnetization states, we increase the magnetic field to $300$\,mT with a $15$\,mT increment. The number of spin objects is counted by a trained neural network based on \textit{yolov8}~\cite{yolov8}.

\begin{figure}[t]
     \centering
     \includegraphics[width=\columnwidth]{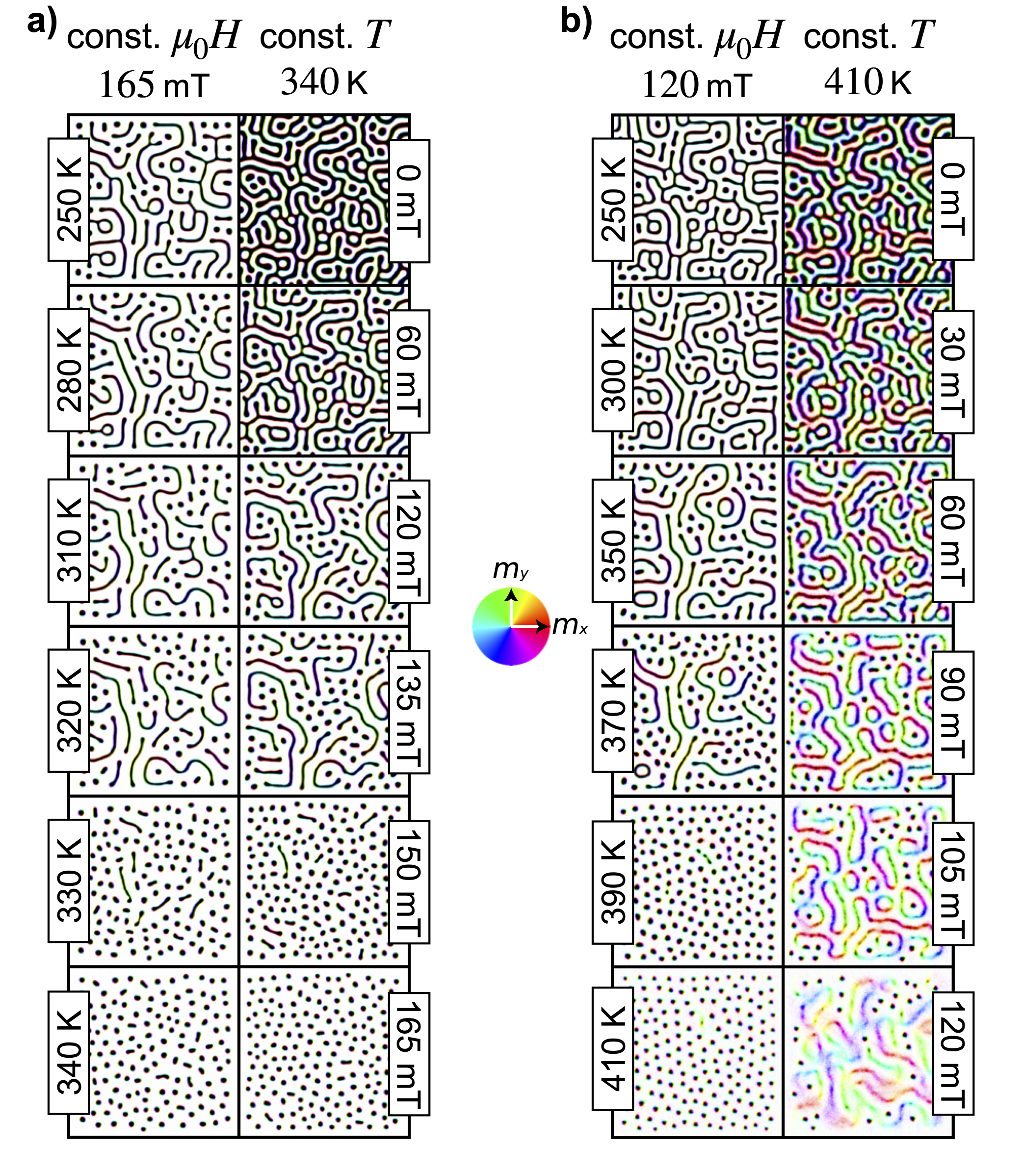}
     \caption{\textbf{Pathway-dependent magnetic spin textures obtained from micromagnetic simulations.} \textbf{a)} Evolution of spin textures toward ($\mu_0H,T$)\,$=$\,(165\,mT, 340\,K), where both paths predict quite similar numbers of BSKs in the final state. \textbf{b)} Evolution of spin textures toward ($\mu_0H,T$)\,$=$\,(120\,mT, 410\,K), where BSKs are only strongly present in case of the constant magnetic field trajectory. Depicted simulation areas are $5.12\times 5.12$\,{\textmu}m$^2$.}
     \label{fig:SimTexture}
\end{figure}

The magnetic field- and temperature-dependent micromagnetic simulations, displayed in Fig.~\ref{fig:Sim}a), reveal a significant path-dependence of the resulting phase diagrams. Remarkably, as in the experiment, the BSK phase extends to higher temperatures for the case of constant fields. The difference of both diagrams shown in Fig.~\ref{fig:Sim}b highlights the fundamental distinctions between the two superficially similar pathways. While red colors indicate more BSKs for the pathway of constant magnetic field, blue colors denote more BSKs for the pathway of constant temperature. Evidently, the pathway of constant magnetic field enhances the stability of BSKs for both higher magnetic fields and higher temperatures (red colored rectangles), qualitatively similar to the experiment. In contrast, the large blue region indicates that less BSKs are created for intermediate temperatures and magnetic fields, i.e., at the transition from stripe domains to BSKs. To obtain a mesoscale view on the different evolution of spin textures for different pathways, snapshots of simulated magnetic spin textures for the ($H,T$)-pathways marked by arrows in Fig.~\ref{fig:Sim}b are presented in Fig.~\ref{fig:SimTexture}. 

We start by discussing the two pathways converging to ($\mu_0H$, $T$)\,$=$\,(165\,mT, 340\,K) in Fig.~\ref{fig:Sim}a, where both paths predict a relatively similar final number of BSKs. Even for a relatively similar final state, the evolution of the two paths to the point ($\mu_0H$, $T$)\,$=$\,(165\,mT, 340\,K) is apparently quite different, see Fig.~\ref{fig:SimTexture}a. For constant field, the system transitions from a moderately dense stripe domain state into a BSK lattice state with slightly lower density and ordering than at for the case of increasing the magnetic field at constant $T$. Here, we observe a transition from a dense stripe domain state into a dense BSK lattice with relatively high ordering. At the depicted final temperature of $T=340$\,K this effect is weak, for most intermediate fields and temperatures it is much stronger, as indicated by the blue colored regions in Fig.~\ref{fig:Sim}b.

Differences in the micromagnetic simulations get much more pronounced studying the trajectories to ($\mu_0H,T$)\,$=$\,(120\,mT,
410\,K), Fig.~\ref{fig:SimTexture}b. Here, we start from initially relatively similar spin textures, which however evolve in drastically different manner. For constant $H$ and increasing $T$, the stripe domain state gradually evolves into a very dense and highly ordered BSK lattice at 410\,K, whereas in case of starting at high $T$ and increasing $H$ nearly no BSK nucleation takes place and a weak stripe domain pattern remains with few interdispersed cylindrical spin objects. In the latter case, the spin structures resemble the domain morphology of a chiral magnet such as Fe$_{1-x}$Co$_x$Si, FeGe, or MnSi, where the helical or conical phases form due to the presence of a bulk DMI~\cite{Uchida2006,Yu2011,Tonomura2012}. In these systems, the magnetization orients oop, and forms chiral skyrmions and lattices, even though they lack the pronounced perpendicular magnetic anisotropy that we can see in the system under investigation. Moreover, magnetization patterns with increased in-plane features are visible. The formation of closed domains appears similar to the magnetic configurations of hopfionic rings~\cite{Zheng2023}. However, after visually investigating their 3D spin configurations, we can conclude that these structures do not have all the characteristics of hopfions or hopfion rings. Nevertheless, given the thickness of our film, we cannot exclude the possibility that ferrimagnetic multilayers (and also alloys) might be inherently hosting meta-stable hopfionic states and that the path to a certain environmental condition is another possible route to stabilize new or unexpected states.

Apart from these specific observations, a comparison between Fig.~\ref{fig:SimTexture}a~and~b clearly highlights that for, e.g., the constant field case, the micromagnetic simulations predict that the final BSK lattice density and ordering strongly depends on the choice of the initial field condition.

\section{Results - Strong perturbation}
\subsection{Phase map in comparison to the weak excitation case}
\begin{figure}[t]
     \centering
     \includegraphics[width=\columnwidth]{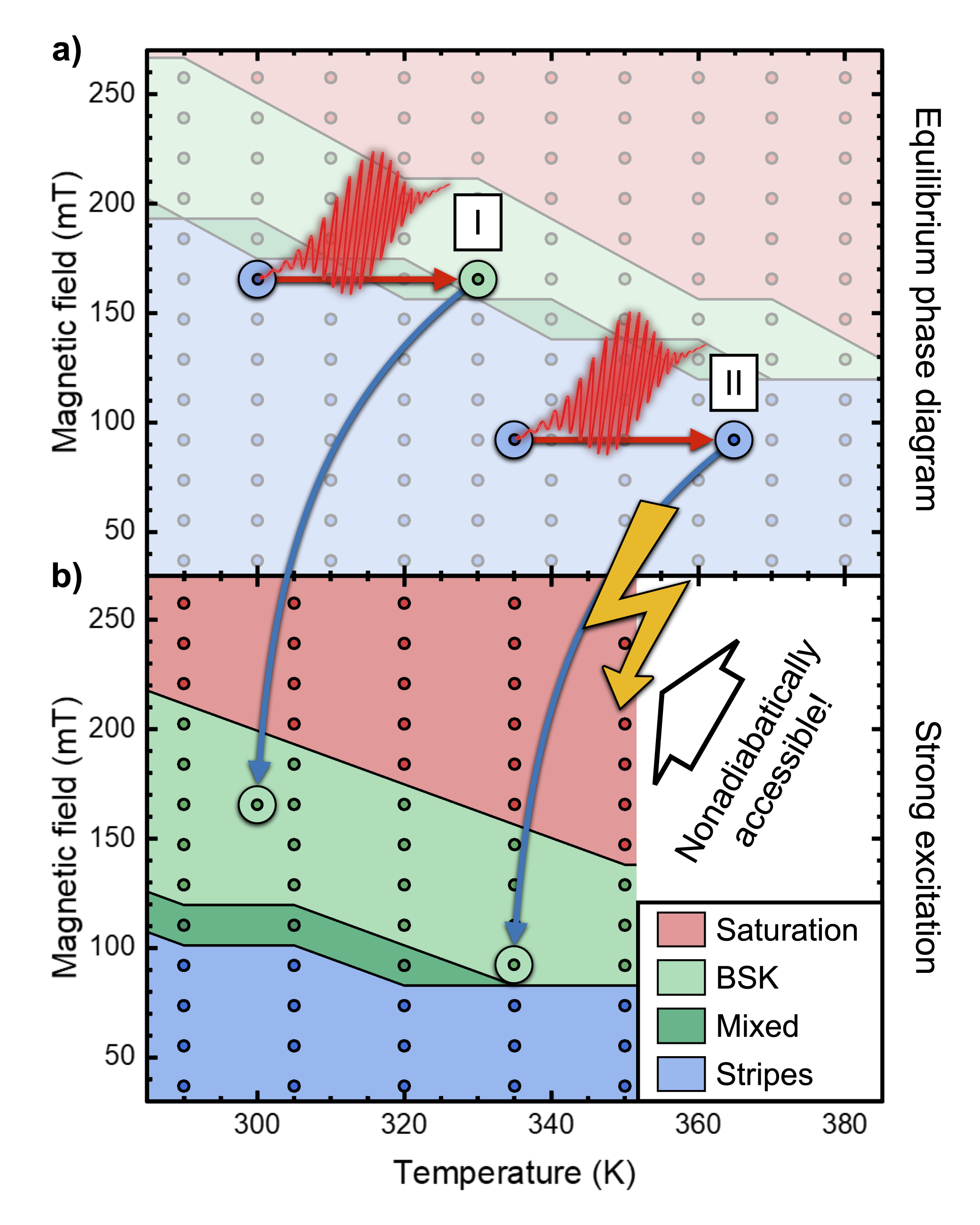}
     \caption{\textbf{Modification of magnetic spin textures using strong optical excitation.} \textbf{a)} Phase diagram of the ground state magnetic spin textures, which is obtained from the weak perturbation measurements presented in Fig.~\ref{fig:BandTsweep}d. \textbf{b)} Strong optical excitation leads to a transformation of the phase diagram. Two pathways from (a) to (b) are qualitatively highlighted as (I) and (II), where only (I) can be explained in the framework of adiabatic sample heating. Color codes are again red for the saturated state, green for the BSK state and blue for the stripe domain state.}
     \label{fig:Modification}
\end{figure}

Previously~\cite{Titze2024a}, we demonstrated that a strong perturbation is capable of modifying the magnetic spin textures, i.e. transforming stripe domains into BSKs and annihilating BSKs. Here, we follow this ansatz and repeated the magnetic field-dependent measurements at various temperatures under strong laser excitation $F=4$\,mJ/cm$^2$. The corresponding phase diagram is depicted in Fig.~\ref{fig:Modification} in comparison to the equilibrium phase diagram, which is obtained from the weak excitation case at constant magnetic fields. This allows us to explore the origin of the modification of magnetic spin textures upon strong laser excitation. 

In accordance to previous measurements at room temperature~\cite{Titze2024a}, the magnetic field regime, in which the BSK lattice phase is stabilized, is shifted to lower magnetic field values, which is attributed to laser-induced BSK nucleation and annihilation processes. Interestingly, a BSK lattice is observed even at temperatures $T\geq 350\,$K, for which it does not appear in the ground state phase diagram for measurements at constant temperature, see Fig.~\ref{fig:BandTsweep}b. However, since laser excitation introduces heat to the sample system, one rather has to compare the phase diagram to the case of heating at constant magnetic fields depicted in Fig.~\ref{fig:BandTsweep}d, which is reprinted in Fig.~\ref{fig:Modification}a. 

Comparing the phase diagram for strong excitation to the weak excitation scenario allows to gain insight into the mechanism behind the modification of magnetic spin textures. Assuming purely adiabatic thermal effects, the strong excitation can be imagined to heat up the system, so that a stable BSK lattice phase is entered (red arrow, scenario I, assumed heating of spin system by 30\,K~\footnote{The sample heating can be estimated by the average heating of the system by the strong excitation derived in Fig.~\ref{fig:HighFluence} of around 23\,K. However, this temperature increase represents only a lower boundary. The value of 30\,K used here coincides with the approximate shift of the BSK to single domain transition temperature on the temperature axis that we observe for comparing the weak and strong excitation cases.}). This heating process is followed by a subsequent thermal quench, which freezes the BSK lattice, similar to what has been reported for chiral magnets~\cite{Berruto2018,Kalin2024}, highlighting that the energy barriers for the creation and annihilation of skyrmions are different. Evidently, the process marked I) can in principle be explained by such a thermal nucleation process.

However, the comparison between both phase diagrams displayed in Fig.~\ref{fig:Modification} reveals also BSK states, e.g., via process II), which clearly contradict a thermal creation process following an equilibrium pathway, indicating that they are created via a nonadiabatic process. In addition, we find that for all temperatures the BSK lattice phase is significantly broader in the case of strong optical excitation, further corroborating the presence of nonadiabatic processes.
  
\subsection{Stability of laser-induced spin textures}
\begin{figure}[t]
     \centering
     \includegraphics[width=\columnwidth]{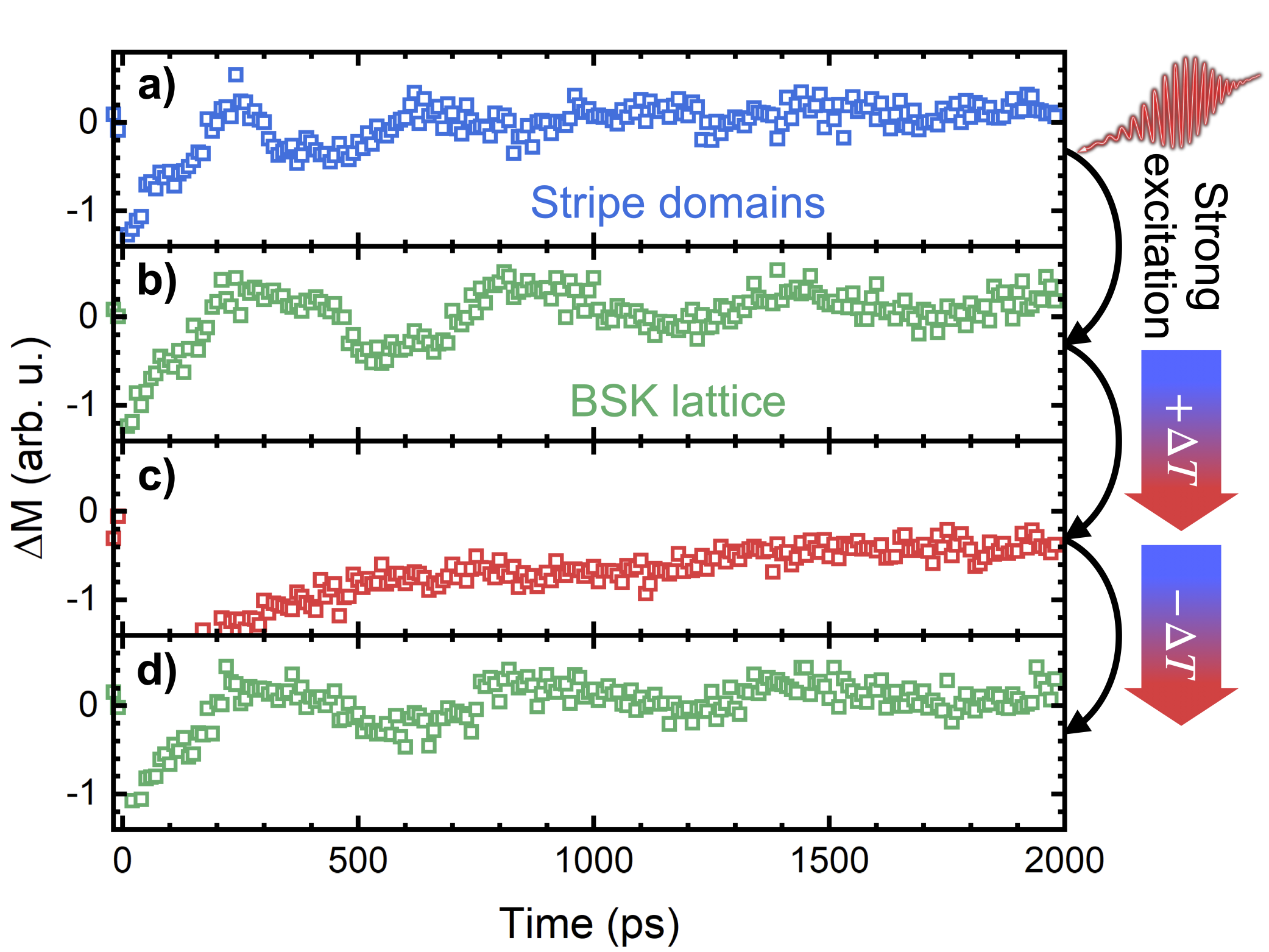}
     \caption{\textbf{Stability of a nucleated BSK phase.} \textbf{a)} Magnetization dynamics obtained from a stripe domain phase at $T=300$\,K and $\mu_0 H=156$\,mT using weak optical excitation ($F=0.6$\,mJ/cm$^2$). \textbf{b)} Magnetization dynamics after the sample has been exposed to strong laser excitation ($F=6$\,mJ/cm$^2$). \textbf{c)} The temperature is increased to $T=360$\,K. \textbf{d)} The temperature is decreased back to $T=300$\,K.}
     \label{fig:BSKstability}
\end{figure}

In Fig.~\ref{fig:BSKstability} we now focus on the temperature-stability of such a laser-induced state. First, we stabilized a stripe domain state at $T=300$\,K by applying an oop magnetic field $\mu_0 H=156$\,mT. Magnetization dynamics in the weak perturbation limit ($F=0.6$\,mJ/cm$^2$) in Fig.~\ref{fig:BSKstability}a) reveal the characteristic stripe domain breathing mode. Exposing the sample to strong optical excitation ($F=6$\,mJ/cm$^2$) leads to the nucleation of the BSK state. Reducing the laser fluence back to the weak perturbation limit, this BSK state remains stable, which is evident from the frequency shift present in Fig.~\ref{fig:BSKstability}b). 

An increase of temperature to $T=360$\,K in Fig.~\ref{fig:BSKstability}c) yields a transient, which at a first glance bears resemble of a transient to the saturated magnetic state. A more detailed analysis reveals both a softening of the breathing mode from $1.7$\,GHz to $1.1$\,GHz and a strong attenuation of its amplitude by a factor of 5. This in turn might indicate that BSKs are mostly annihilated due to the temperature increase, i.e., the sample is close to but not yet in the saturated state. However, subsequent decrease of the temperature to $T=300$\,K in Fig.~\ref{fig:BSKstability}d) reveals quite similar magnetization dynamics as before the heating. We therefore rule out temperature-induced BSK annihilation as a significant cause of the decrease in amplitude. Instead, we attribute it to a reduced amplitude of the BSK breathing mode. Breathing is initiated by an expansion of the BSK cores upon laser excitation due to a thermally-activated change of the magnetic anisotropy, i.e., the effective magnetic field~\cite{Titze2024b}. If the BSK core size is strongly reduced due to the increase in base temperature, further laser-induced changes yield only small effects. These result in a strong reduction of the observed breathing mode amplitude, while the softening is induced by the static temperature increase (see Appendix, Fig.~\ref{fig:Amplitudes}c). 

\section{Discussion}
Our results in case of weak optical excitation strongly indicate that the path to a certain point in the ($H, T$)-phase diagram allows choosing different and potentially new magnetic configurations. The micromagnetic simulations predict that enclosed circular domains could be stabilized by in-field heating, which were not stable under a different protocol. It is noteworthy that alternating protocols for heating and applying a combination of ip/oop fields with concurrent heating/cooling has been demonstrated to allow the formation of antiskyrmion lattices in Heusler compounds, or skyrmion lattices in 2D materials~\cite{Jena2020,Powalla2023,Jena2024}. Similarly, our experimental and computational results suggest that the path itself is an important degree of freedom for reaching novel metastable states. Furthermore, the micromagnetic simulations clearly indicate, that the density of spin objects within a given phase can be controlled by the chosen creation path. 

For strong excitation, we observe laser-nucleation of BSK outside the equilibrium phase diagram. In general, laser-induced nucleation processes of skyrmions are discussed in the literature to fall into two categories: (i) thermal, quasistatic excitation into an equilibrium skyrmion phase at elevated $T$ with a subsequent quench into a metastable state~\cite{Berruto2018,Kalin2024}, and (ii) excitation of the system into a strongly nonequilibrium fluctuation state leading to the creation of topological spin defects and nucleation of skyrmions in the cooling of the system~\cite{Buettner2021,Khela2023, Truc2023}. As evidenced by Fig.~\ref{fig:Modification}, the creation of a BSK lattice in our case cannot unequivocally fall into the former category (process I), but necessarily has to involve the creation of topological defects in the system, as process II cannot be explained in a quasistatic thermal picture.

Indeed, the creation of topological defects is shown to play a major role in both the annihilation~\cite{Li2020, Birch2021}, but also the creation of skyrmions~\cite{Li2020}. So-called hedgehogs or Bloch points, topological magnetic singularities or monopoles, are a necessary ingredient to allow for winding or unwinding of skyrmionic textures. In a recent publication~\cite{Li2020}, Li and coworkers discuss how nucleation of skyrmions in a decreasing magnetic field is possible at an artifically introduced notch defect in a magnetic structure. The local modification of the system's demagnetization field at the notch modifies the magnetostatic energy density, enabling the formation of a Bloch point at a certain critical field. This Bloch point then propagates through the film, leading to the creation of a skyrmion. 

\begin{figure}[t]
     \centering
     \includegraphics[width=\columnwidth]{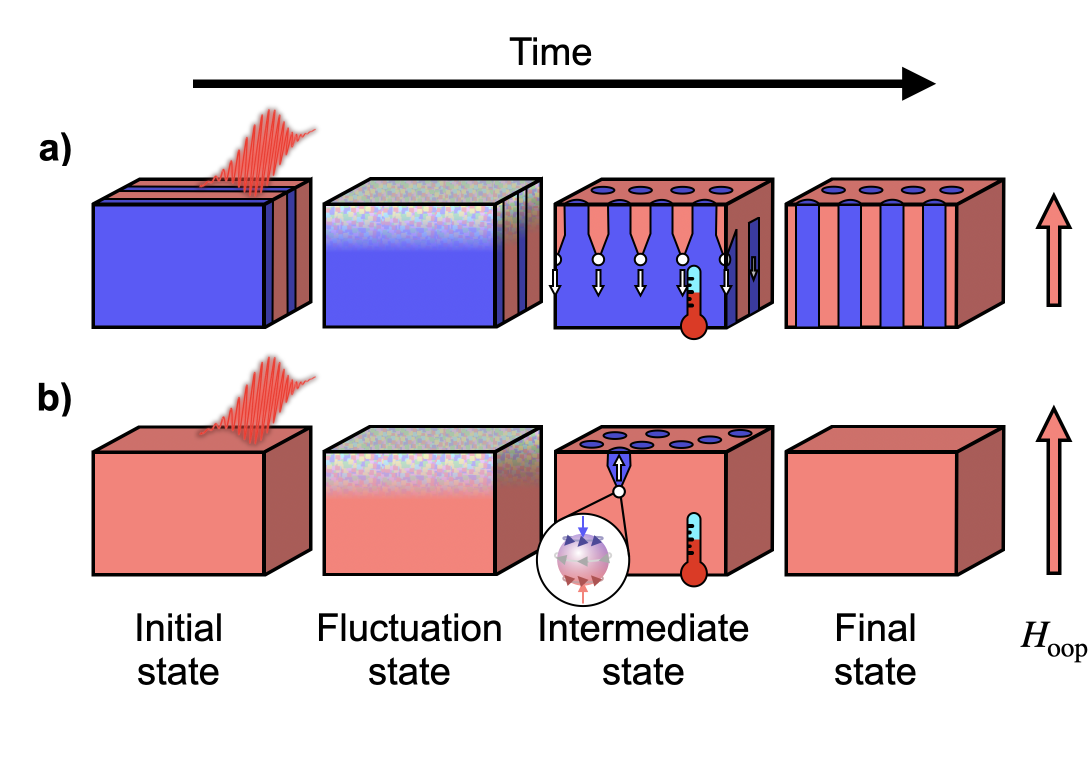}
     \caption{\textbf{Schematic depiction of the mechanism of BSK nucleation.} Magnetic domains are colored in red and blue corresponding to up and down orientation of the magnetization. Optical excitation of the initial state leads to a fluctuation state of the topmost layers, which transitions to the final state via a hot intermediate state. \textbf{a)}~Starting from a stripe domain state, Bloch points (white dots) are introduced that travel into the film and give rise to a \textsc{BSK} lattice retaining the pattern and chirality of the original stripe domain state~\cite{Titze2024a}. \textbf{b)}~Starting from the single-domain state, Bloch points are introduced, which are unstable in the opposing magnetic field and the systems collapses back into a single-domain state.}
     \label{fig:Mechanism}
\end{figure}

In our case, the local surface excitation of the 120\,nm thick Fe/Gd multilayer in the stripe domain state has two effects (see Fig.~\ref{fig:Mechanism}a). Close to the sample surface, within the first 10--20\,nm, the magnetization is strongly or partly even totally quenched by ultrafast demagnetization. This may allow to locally create a fluctuation state similar to the results reported in Ref.~\cite{Buettner2021}. Second, and possibly even more important, the demagnetizing field is strongly reduced by the quenched magnetization in the surface region. Thereby, we achieve a spatially nonuniform magnetization state, similar to the artifically created notch discussed in Ref.~\cite{Li2020} or as obtained by using an magnetic force microscope tip in Ref.~\cite{Zhang2018a}. Both the fluctuation state as well as the nonuniform magnetization may thus lead to the creation of Bloch points. We expect these to appear in between the domain walls of the stripe domains, as the spatially rapidly changing spin orientation in these regions is already close to such a state. The splitting of stripe domains in this way then leads to the creation of cylindrical spin objects by the extension of the energetically favorable spin regions with orientation parallel to the field by the downwards moving Bloch points (see intermediate state in Fig.~\ref{fig:Mechanism}a).

In contrast, any fluctuation state or topological defects induced in the fully saturated state cannot induce a stable nucleation of skyrmions according to our experimental observations here and in Ref.~\cite{Titze2024a}. Even if smalls regions of reversed magnetization are present at random positions after the initial fluctuation state, these seem to be energetically unfavorable and the corresponding Bloch point will move to the surface to be annihilated (see Fig.~\ref{fig:Mechanism}b). 

The explanations above clearly rely on the local inhomogeneity of the system in form of spin textures (stripe domains) and also localized heating to create a modified surface magnetization. In fact, localized heating has been described theoretically and experimentally as a mechanism for skyrmion creation~\cite{Koshibae2014, Liu2024,Wang2024}. Our explanation also clarifies why a difference exists between the quasistatic ($H,T$)-phase diagram for constant $H$ and the ($H,T$)-phase diagram for strong excitation. In the quasistatic or adiabatic case, all relevant system parameters change slowly and homogeneously throughout the film, allowing for the creation of BSKs only for a suitable global magnetic energy landscape. In contrast, femtosecond optical excitation creates a sudden strong temperature increase quenching $M$ locally, while a large part of the film remains initially at a higher magnetization and lower temperature. This inhomogeneous situation then allows for the BSK growth in regions of the phase diagram, where this is normally not possible. They are ultimately stable in extended regions of the ($H,T$)-phase diagram after global system equilibration, as the annihilation barrier is typically different from the nucleation barrier~\cite{Zhang2018b,Li2020}. This difference between the barriers for creation and annihilation is also readily seen from Fig.~\ref{fig:BSKstability}. Here, the BSKs persist even in case of a reduced pump fluence, where any static heating by the strong laser excitation is strongly diminished. We note that the obtained nonequilibrium phase diagram in Fig.~\ref{fig:Modification}b might not be necessarily nonthermal as such, as temperature may be locally well defined everywhere after the first few picoseconds and all subsystems like electrons, phonons and spins may locally be both in internal equilibrium and in equilibrium with each other. Instead, the fact that local gradients exist in $T$, $M(T)$, the topology, and all derived quantities in the system like demagnetizing field and anisotropy may very likely be the main driving factor for the observed behavior. The driving nonequilibrium then does not originate in the time-domain, but in space.

\section{Conclusion}
In summary, we have shown that quasistatic variation of both temperature and magnetic field allows for adjusting the magnetic spin texture in Fe/Gd multilayers. However, the final state does not only depend on the values of temperature and magnetic field, but also on the path travelled within the ($H,\,T$)-phase diagram, highlighting the intrinsic memory effects in magnetic materials. The stability range of the BSK phase can be drastically increased by sweeping temperature instead of magnetic field, which is corroborated by micromagnetic simulations showing the intrinsic differences in domain pattern between the different pathways. 
In addition, we employed strong femtosecond optical excitation to modify the magnetic spin textures. We observe a drastic change in the stability range of the BSK phase extending it to much lower magnetic fields. Taking thermal effects into consideration reveals that this BSK nucleation is at least partially driven by nonadiabatic processes, which allow for stabilization of nonequilibrium BSK phases outside the phase space of the equilibrium ($H,\,T$)-phase diagram. We expect local, strong variations in $M$ and derived quantities to be the main contributing factor allowing for the creation of laser-nucleated BSKs, meaning that the nonequilibrium is predominantly of spatial and not temporal nature. 

In total, our study elaborates how different pathways can be used to greatly extend the range of ($H,\,T$) combinations, in which BSK lattices can be stabilized in Fe/Gd multilayers. Furthermore, we highlight that in samples thicker than the laser penetration depth spatial nonequilibrium may be a main driving factor for laser-induced BSK nucleation. Thereby, we provide guidelines for stabilization mechanisms of magnetic spin textures, laying necessary groundwork toward future texture-based spintronic and magnonic devices.  
\begin{acknowledgments}
T.T.\,and D.S.\,gratefully acknowledge funding by the German Research Foundation (DFG, Grant No.\,217133147 (SFB1073, project A02).T.S.\,and M.A.\,gratefully acknowledge funding by the DFG, Grant No.\,507821284.  S.K.\,and D.Su.\,acknowledge the Austrian Science Fund (FWF) for support through Grant No.\,I\,6267 (CHIRALSPIN). S.K.\,thanks the Vienna Doctoral School in Physics for funding
the Mobility Fellowship. S.K.\,acknowledges funding from the European Research Council (ERC) under the European Union’s Horizon 2020 research and innovation programme, Grant Agreement No.\,101001290 (3DNANOMAG). We acknowledge Vienna Scientific Cluster (VSC) for awarding this project access to the LEONARDO supercomputer, owned by the EuroHPC Joint Undertaking, hosted by CINECA (Italy) and the LEONARDO consortium.
\end{acknowledgments}

\subsection*{Author Contributions}
S.\,Mathias and D.\,Steil conceived the project. T.\,Titze and M.\,Matthies performed
the time-resolved Kerr effect studies and analyzed the time-resolved magnetization data
under the supervision of S.\,Mathias and D.\,Steil. Sample growth and characterization was performed by T.\,Schmidt under the supervision of M.\,Albrecht. S.\,Koraltan performed the micromagnetic simulations under the supervision of D.\,Suess. The simulations were evaluated by S.\,Koraltan and T.\,Titze with support from D.\,Steil. T.\,Titze, S.\,Koraltan and D.\,Steil wrote the manuscript with input from all authors.

\section{Appendix}
\subsection{T-dependence of mode dispersion}\label{AppB:Dispersion}
\begin{figure}[ht]
     \centering
     \includegraphics[width=\columnwidth]{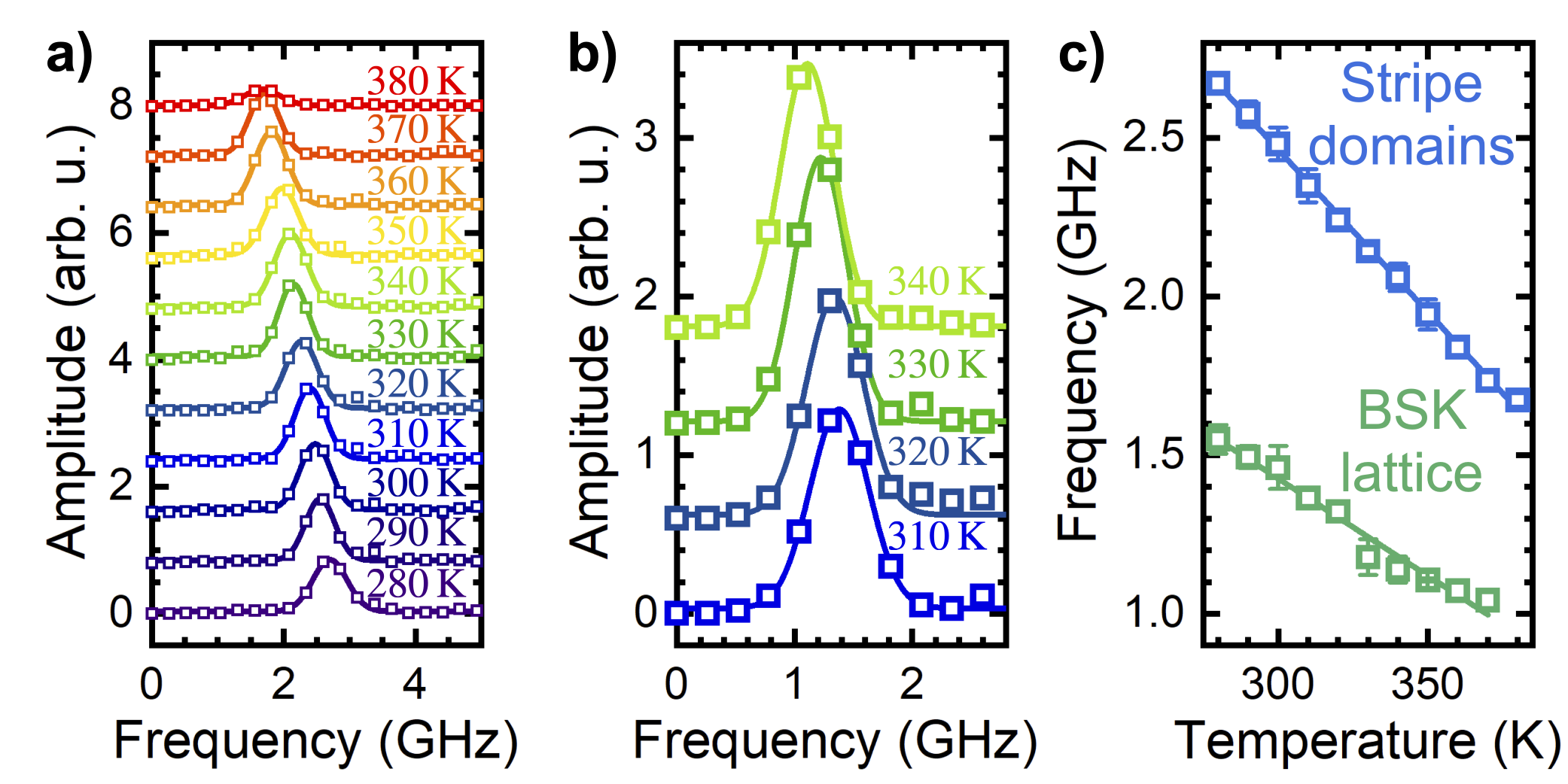}
     \caption{\textbf{Temperature-dependent analysis of the breathing modes.} Temperature-dependent Fourier amplitudes for \textbf{a)} stripe domains ($\mu_0 H=110$\,mT) and \textbf{b)} BSKs ($\mu_0 H=184$\,mT). \textbf{c)} Breathing mode frequencies averaged over all magnetic fields extracted from a Gaussian fit.}
     \label{fig:Amplitudes}
\end{figure}

Fig.~\ref{fig:Amplitudes}a,b depict the Fourier-amplitudes for both the stripe domain ($\mu_0 H=110$\,mT) and BSK lattice ($\mu_0 H=184$\,mT) phase for different temperatures. We find that in both cases the breathing mode frequency decreases with increasing temperature. A Gaussian fit to the Fourier-amplitude reveals a linear behaviour of the temperature-dependent breathing mode frequency, see Fig.~\ref{fig:Amplitudes}c. Here, we exploit that the breathing mode frequencies do barely change with the external magnetic field, see Fig.~\ref{fig:BandTsweep}a,c. Therefore, for each temperature multiple data points are averaged, error bars are the corresponding standard deviation. Interestingly, the difference in slope, considering the breathing mode of stripe domains and the BSK lattice, is related to their difference in frequency, i.e., they share almost the same zero crossing as we find $T_{0,\,\mathrm{bsk}}=533\pm 14$\,K and $T_{0,\,\mathrm{st}}=540\pm 4$\,K. Furthermore, the decrease in frequency perfectly scales with the temperature-induced reduction of $M_S$ and the effective magnetic anisotropy~\cite{Montoya2017b}.

\subsection{Average T-increase for strong perturbation}\label{AppC:StrongPerturbation}
\begin{figure}[h!]
     \centering
     \includegraphics[width=\columnwidth]{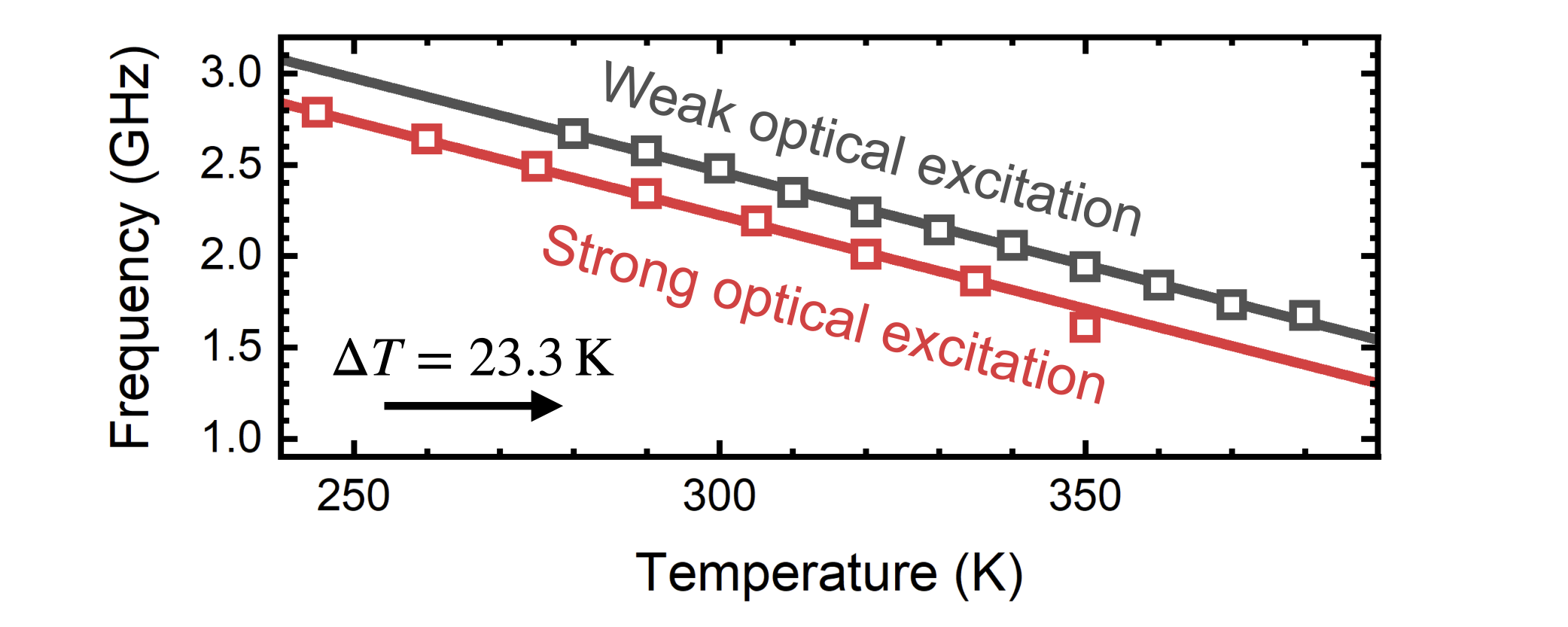}
     \caption{\textbf{Temperature-dependent breathing mode frequency for the stripe domain phase upon weak and strong optical excitation.} The temperature of the strong excitation data has to be shifted by $\Delta T=23.3$\,K in order to obtain perfect agreement with the weak excitation data.}
     \label{fig:HighFluence}
\end{figure}
\newpage
The temperature-dependent breathing mode frequency of stripe domains, considering strong optical excitation, is depicted in Fig.~\ref{fig:HighFluence}. It allows us to extract an average laser-induced temperature increase within our temporal observation window by mapping the data for strong excitation onto the data for weak excitation. We achieve perfect agreement by introducing a temperature shift $\Delta T=23.3$\,K. 

%

\end{document}